%% file: faraday.tex
\documentclass[aps,pra,reprint,groupedaddress,floatfix]{revtex4-1}

\input{preamble}

\begin{document}

\title{Magic-wavelength Faraday probe measures spin continuously and without light shifts}

\author{M.~Jasperse}
\author{M.\,J.~Kewming}
\author{S.\,N.~Fischer}
\author{P.~Pakkiam}
\author{R.\,P.~Anderson}
\author{L.\,D.~Turner}\email[]{lincoln.turner@monash.edu}
\affiliation{School of Physics \& Astronomy, Monash University, Victoria 3800, Australia.}

\date{\today}

\begin{abstract}
  We describe a dispersive Faraday optical probe of atomic spin which performs a weak measurement of spin projection of a quantum gas continuously for more than one second.
  To date focusing bright far-off-resonance probes onto quantum gases has proved invasive, due to strong scalar and vector light shifts exerting dipole and Stern-Gerlach forces.
  We show that tuning the probe near the magic-zero wavelength at $\unit[790]{nm}$ between the fine-structure doublet of \Rb cancels the scalar light shift, and careful control of polarization eliminates the vector light shift.
  Faraday rotations due to each fine-structure line reinforce at this wavelength, enhancing the signal-to-noise ratio for a fixed rate of probe-induced decoherence.
  Using this minimally-invasive spin probe we perform microscale atomic magnetometry at high temporal resolution.
  Spectrogram analysis of the Larmor precession signal of a single spinor Bose-Einstein condensate measures a time-varying magnetic field strength with $\unit[1]{\upmu G}$ accuracy every $\unit[5]{ms}$; or equivalently makes $>200$ successive measurements each at $\unit[10]{pT/\sqrt{Hz}}$ sensitivity.
\end{abstract}


\maketitle

\section{Introduction}
\label{sec:introduction}
Dispersive probes of quantum systems deliver temporally-rich data, expose fluctuating and critical processes, maximally exploit long coherence times and enable feedback control. 
The Faraday light-matter interface couples atomic spin via the off-resonant vector electric dipole interaction to optical polarization, which is readily measured at the quantum limit.
Faraday measurements have opened new perspectives in quantum metrology~\cite{budker_optical_2007}, quantum information~\cite{muschik_quantum_2011}, non-linear mean-field~\cite{liu_quantum_2009} and many-body~\cite{behbood_generation_2014} systems; and have potential for probing strongly-correlated systems~\cite{eckert_quantum_2008}.
The Faraday interface has been applied to progressively colder systems: magneto-optical traps~\cite{isayama_observation_1999}, dark- ~\cite{fatemi_spatially_2010} and bright-optical dipole traps~\cite{kubasik_polarization_2009,behbood_real-time_2013,behbood_feedback_2013,sewell_certified_2013,behbood_generation_2014,colangelo_simultaneous_2017}, and Bose-Einstein condensates (BEC)~\cite{liu_quantum_2009,liu_number_2009}.
However, the classical backaction of the Faraday probe perturbs atomic motional and spin degrees of freedom, limiting and confounding measurements of emergent phenomena or weak external fields at ultracold temperatures.
Here we explore theoretically and experimentally a magic-wavelength Faraday probe which minimizes classical backaction, and so enables continuous measurements of a coherent spinor quantum gas beyond one second. 

\subsection{The Faraday light-matter interface}
\label{sec:interface}
The resonant Faraday effect is the enhanced rotation of light polarization by an atomic vapor in the wings of absorption lines, first observed in 1898~\cite{macaluso_1898a,*macaluso_1898b,*macaluso_1898c}.
As a spectroscopic tool, the effect has been exploited for ultranarrow optical filters~\cite{ohman_apparatus_1956,kiefer_faraday_2014}, ultrasensitive spectroscopy~\cite{brumfield_faraday_2012}, laser stabilization~\cite{marchant_offresonance_2011} and optical frequency standards~\cite{zhuang_active_2014}.
Resonant Faraday probing of spins undergoing Larmor precession results in the polarization angle of transmitted light oscillating at the Larmor frequency.
Such polarization modulation is readily photodetected with a balanced polarimeter, and given knowledge of the gyromagnetic ratio yields calibration-free `optical magnetometry', widely implemented in warm atomic vapors~\cite{budker_book_2013}.
Faraday probes enable quantum state tomography of atomic gases, measuring a single spin projection in the laboratory frame while the spin state is evolved by applied radiofrequency and microwave fields~\cite{smith_efficient_2006,riofrio_quantum_2011}, or by spin-mixing interactions in a degenerate spinor gas~\cite{liu_quantum_2009}.

When the coupling between the light polarization and spin polarization is strong, the quantum backaction of the light rotating the spins becomes important, and the Faraday interaction is considered a `quantum light-matter interface'~\cite{hammerer_quantum_2010}, rather than an optical probe of an unperturbed atomic sample. 
Used with warm atomic vapors, this `Faraday interface' has squeezed spins~\cite{kuzmich_generation_2000}, squeezed light~\cite{mMkhailov_lowfrequency_2008}, entangled states of collective atomic ensembles~\cite{julsgaard_experimental_2001}, encoded light into quantum memories~\cite{julsgaard_experimental_2004} and teleported states of light to atoms~\cite{sherson_quantum_2006}.
Applied to ultracold (but not degenerate) atoms~\cite{kubasik_polarization_2009}, the Faraday interface has created macroscopic singlet states~\cite{behbood_generation_2014}, squeezed two spin projections simultaneously~\cite{colangelo_simultaneous_2017}, cooled by feedback~\cite{behbood_feedback_2013} and made macroscopic quantum non-demolition measurements~\cite{sewell_certified_2013}.

\subsection{Choice of detuning}
\label{sec:detuning}
The parameters of the Faraday interface itself -- in both strong- and weak-coupling regime -- have been studied including the optimal spatial mode and its effects on quantum noise~\cite{sorenson_three-dimensional_2008} and backaction~\cite{baragiola_three-dimensional_2014}; the temporal mode for pulsed probes~\cite{vasilakis_stroboscopic_2011}; and the effect of polarization on tensor light shifts~\cite{smith_tensor_2004}.
The effect of probe detuning $\Delta$, however, has only been studied in terms of the off-resonant photon scattering rate,
i.e. the incoherent component of the backaction of the probe on the atoms. 
In the case of a probe far-detuned from an isolated spectral line the Faraday rotation angle is dispersive and falls off as $1/\Delta$, yielding a polarimeter signal proportional to $I_0/\Delta$ for probe intensity $I_0$.
Because the photon quantum noise of this probe is proportional to $\sqrt{I_0}$, the photon-limited signal-to-noise ratio (SNR) for a Faraday measurement scales as $\sqrt{I_0}/\Delta$ or equivalently as $\sqrt{\gamma_s}$, where $\gamma_s$ is the off-resonant photon scattering rate~\cite{smith_faraday_2003}.
This two-level Faraday measurement model has been extended to more realistic atomic structure, including all hyperfine lines of a \emph{single} fine-structure transition~\cite{vasilyev_quantum_2012}.
These models conclude that the Faraday SNR depends strictly on scattering rate, with detuning a free parameter; e.g. one can choose a low-intensity probe tuned close to resonance or a brighter beam at larger detuning.

The experimenter is soon caught in a dilemma.
A practical choice is to use the brightest possible beam, to overwhelm the technical noise in the photodetector, and sufficiently detuned to reduce the scattering rate to the maximum permissible level.
The \emph{coherent} backaction of this bright, off-resonant beam includes a scalar light shift of the atomic eigenstates, which scales as $I_0/\Delta \propto \Delta \gamma_s$, and for a spatially-varying probe intensity this acts as a dipole trap.
Herein lies the apparent dilemma: the choice of bright beams (at large detuning) comes at the cost of an undesired dipole potential.
The alternative of dim probe beams (at small detunings) minimizes the dipole force but poses the challenge of low-noise wideband photodetection.
In some cases dipole forces are not a problem, e.g. warm vapors whose temperature is much larger than the scalar light shift, cold atoms in deep optical lattices~\cite{smith_faraday_2003}, but experiments with ultracold samples that are weakly-confined (i.e. not in lattices) are acutely affected by probe dipole forces.

\subsection{Backaction dipole forces}
\label{sec:dipoleforces}
There are several approaches for reducing the dipole force effects arising from the scalar coherent backaction of the Faraday probe.
A uniform intensity probe exerts no dipole force: making the probe beam very large compared to the sample approaches this limit.
The fraction of the incident probe beam that interacts with the atoms is then very small; in Ref.~\cite{liu_quantum_2009} a $\unit[1]{mm}$ waist probe illuminated a Bose-Einstein condensate (BEC) of Thomas-Fermi radius less than $\unit[10]{\upmu m}$, meaning that of the intense $\unit[50]{mW}$ incident beam less than $\unit[5]{\upmu W}$ arrived at the polarimeter.
A possible alternative, not attempted to our knowledge, is to shape the beam into a flat-topped profile with specialized diffractive or refractive optics.
Superposing a second `probe' beam detuned the other side of resonance can cancel the scalar (and potentially tensor) light shifts~\cite{montano_quantum_2015}, at the expense of additional complexity and reduced SNR due to the additional off-resonant scattering (unless the second beam is also photodetected).

Here we show that it is possible to perform Faraday detection of an alkali ground-state spin with zero scalar light shift backaction on the atoms, using a single-frequency laser beam focused to a Gaussian waist matched to the atomic sample.
We exploit the fine structure of the alkali principal series ($^{2}\text{S} \rightarrow ^{2}\text{P}$) transitions, tuning our Faraday probe near a `magic-zero' or `tune-out' wavelength of $\unit[790.018]{nm}$~\cite{lamporesi_scattering_2010,*leonard_high-precision_2015,*leonard_erratum_2017}, the line-strength-weighted midpoint of the doublet.
At this wavelength the scalar light shift vanishes~\cite{leblanc_species-specific_2007} but, critically, the Faraday rotation contributions from the D1 and D2 lines reinforce.
While operating at the magic-zero wavelength requires very intense beams, we show that this is achieved with reasonable optical powers that are near optimal for broad bandwidth shot-noise limited photodetection.

\subsection{Outline}
\label{sec:outline}
In \refsec{sec:background_theory} we consider the spherical decomposition of the dipole interaction in the far detuned limit.
We discuss the requirements for nulling induced light shifts of the Faraday probe and consider the location several magic-zero wavelengths.
Comparative expressions for the SNR are derived in three different regimes, from which we can identify an optimal choice of magic-zero wavelength for a minimally perturbative shot-noise limited Faraday measurement.
In \refsec{sec:apparatus} we outline the technical details of our spinor quantum gas and Faraday measurement apparatus. 
Finally in \refsec{sec:sig_process} we describe the signal processing and optimization of our Faraday measurement via the cancellation of probe-induced light shifts and suppression of ambient magnetic field gradients. 
In this analysis we introduce the use of spectrograms for precisely evaluating the Larmor frequency, SNR, and for characterizing the amplitude and frequency modulation created by ambient magnetic fields gradients and noise, and spin-dependent collision dynamics. 
\begin{figure}
    \centering
    \includegraphics[width=1.00\columnwidth]{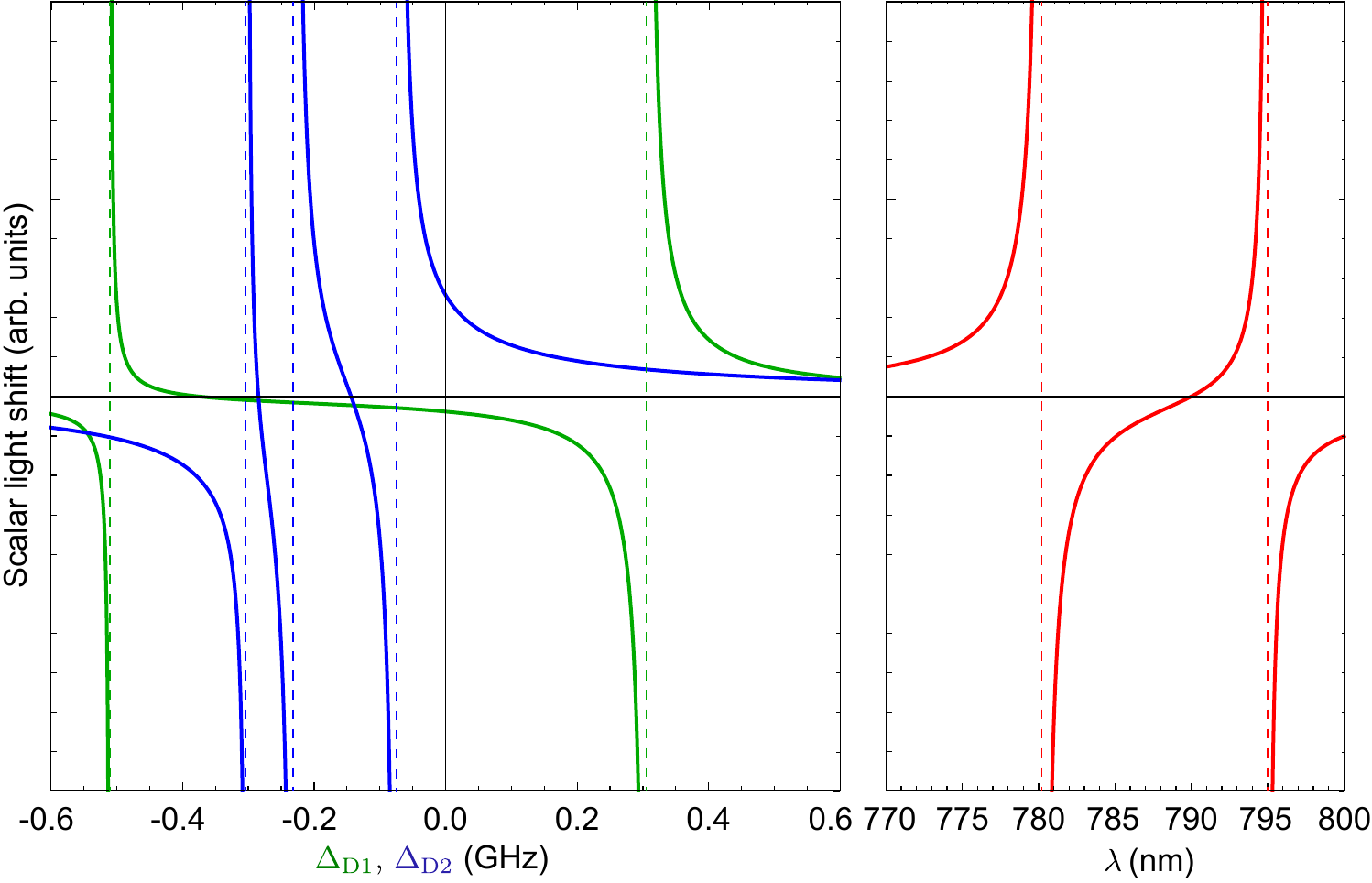}
    \caption{
        \label{fig:tuneout}
        Scalar light shift (arising from $\H_\text{int}^{(0)}$) for \Rb atoms in $F=1$ near the D1 (green) and D2 (blue) lines, and between the two lines (red), showing the detunings relative to respective fine structure transitions (left) and wavelengths (right) at which the scalar interaction vanishes.
        }
\end{figure}

\section{Background Theory}
\label{sec:background_theory}
\subsection{Atom-light interaction}
\label{sec:theory}
In the far-detuned limit, the atom-light dipole interaction is well-described by an effective polarizability Hamiltonian that describes Raman transitions between the ground states. A spherical tensor decomposition allows this interaction Hamiltonian to be written in terms of the probe beam Stokes operators $\op{S}_i$ and the atomic spin operators $\op{F}_i$~\cite{geremia_tensor_2006,stockton_continuous_2007,deutsch_quantum-state_1998}.
Following~\cite{geremia_tensor_2006}, the scalar, vector and tensor contributions can be separated as
\begin{align}
\H_\text{int} &= \H_\text{int}^{(0)} + \H_\text{int}^{(1)} + \H_\text{int}^{(2)} \label{eq:Hint}\\
\H_\text{int}^{(0)} &= g \sum_{J'F'} \frac{\alpha^{(0)}_{J'F'}}{\Delta_{J'F'}} \frac23 \op{S}_0 \label{eq:H0} \\
\H_\text{int}^{(1)} &= g \sum_{J'F'} \frac{\alpha^{(1)}_{J'F'}}{\Delta_{J'F'}} \op{S}_z \op{F}_z\label{eq:H1}\\
\H_\text{int}^{(2)} &= g \sum_{J'F'} \frac{\alpha^{(2)}_{J'F'}}{\Delta_{J'F'}} (\op{S}_x(\op{F}_x^2-\op{F}_y^2) \notag\\[-1em]
					&\qquad\qquad\qquad\quad + \op{S}_y(\op{F}_x \op{F}_y + \op{F}_y \op{F}_x) \notag\\
					&\qquad\qquad\qquad\quad + \op{S}_0 (3 \op{F}_z^2 - F(F+1))/3) \label{eq:H2},
\end{align}
where $g=\omega/2\epsilon_0 V$, $\omega$ is the probe frequency, $V$ is the quantization volume, and $\Delta_{J'F'}$ is the probe detuning above resonance.
The sums is over all excited states $\ket{J'F'}$ permitted by dipole transitions from the ground state $\ket{JF}$, with the coupling strengths $\alpha^{(i)}_{J'F'}$ in this convention defined in App.~(\ref{app:coeff}).

Unlike similar treatments, the explicit $J'$ dependence allows consideration of large detunings where multiple fine-structure transitions contribute.
In particular, we consider the net effect of the D1 ($J'=\frac12$) and D2 ($J'=\frac32$) transitions on the $F=1$ hyperfine states of \Rb.

The scalar interaction $\H_\text{int}^{(0)}$ gives rise to a state-independent energy shift, which can be eliminated by tuning the probe to a `magic-zero' or `tune-out' wavelength~\cite{holmgrem_magic_2012,leblanc_species-specific_2007,arora_tune-out_2011}.
Magic-zero wavelengths exist between each pair of adjacent resonances: at specific detunings within the hyperfine manifold, or between the fine structure transitions (\reffig{fig:tuneout}).
Tuning the probe to one of these wavelengths prevents any dipole forces arising from the spatially-inhomogeneous spatial profile, which is beneficial for long interrogation times.

The vector contribution $\H_\text{int}^{(1)}$ represents a coupling between the Stokes vector of the light and the atomic spin.
A perfectly linearly polarized probe has $\expect{\op{S}_z}=0$ and results in no vector light shift for the atoms as it causes no evolution of the atomic spin operator $\op{\vec{F}}$.

However, spin polarized atoms will cause evolution of the Stokes vector as the probe light propagates through.
This corresponds to rotation of the probe polarization by an angle
$\varphi_z = {\varphi_{0}} \expect{\op{F}_z} / {\xi}_f$, where
\begin{align} 
\varphi_{0} &= 
\frac{\pi\alpha_0\tilde{\rho}}{\epsilon_0 \hbar \lambda},\quad
\frac1{{\xi}_f} = \sum_{J'F'} \frac{\alpha^{(1)}_{J'F'}}{\alpha_{0}\Delta_{J'F'}},
\label{eq:fara_rot0}
\end{align}
$\tilde{\rho}$ is the atomic column density, and $\alpha_0$ is the polarizability constant as derived in App.~(\ref{app:coeff}).
We term $\xi_f$ the ``coherent weighted detuning'', which accounts for the different coupling strengths of different atomic transitions.
The term $\varphi_0$ can alternately be expressed in terms of the on-resonant optical-depth (OD) of any transition $\ket{J}\rightarrow\ket{J'}$ as~\footnote{Typically this result is simplified by taking $\lambda\approx\lambda_{J'}$.}
\begin{equation}
\varphi_{0} = \left(\frac{\lambda_{J'}}{\lambda} \frac{\Gamma_{J'}}{4}\right) \text{OD},
\end{equation}
where $\text{OD} = \tilde{\rho}\,\sigma_{J'}$ and $\sigma_{J'} = 3\lambda_{J'}^2/2\pi$ is the resonant cross-section.

The tensor interaction $\H_\text{int}^{(2)}$ results in complicated evolution of the atomic spin~\cite{smith_tensor_2004}.
For detunings much larger than the excited state hyperfine splittings, it scales as $\mathcal{O}(F^2\gamma_s)$ where $\gamma_s$ is the scattering rate.
The constant of proportionality is order unity~\cite{deutsch_quantum_2010} and is negligible in $F=1$ for measurement times less than the scattering lifetime $1/\gamma_s$.
Although this can be eliminated in pulsed measurement using dynamical decoupling~\cite{koschorreck_dynamical-decoupling_2010}, for continuous measurement this interaction vanishes when the probe polarization is oriented at $\arctan(\sqrt2)=54.7^\circ$ with respect to the external magnetic field~\cite{smith_tensor_2004}.

Choice of the probe polarization and detuning therefore allows the undesirable terms of the polarizability Hamiltonian to be eliminated, leaving only the Faraday interaction that probes the atomic spin state.
This is significant as the spatially-varying probe beam intensity profile would otherwise cause spatially-dependent light shifts that would dephase the atomic spins and result in $\expect{\op{\vec{F}}}\rightarrow\vec{0}$.

\subsection{Signal-to-noise optimization}
\label{ssec:snr_optimization}
A balanced polarimeter splits the probe beam into its polarization components, with respective intensities 
\begin{equation}
I_\pm = I_0\cos^2(\tfrac\pi4 \pm \varphi_z)\Rightarrow I_+-I_-\approx 2I_0\varphi_z \text{ for } \varphi_z\ll1.
\end{equation}
Noting that both the probe beam intensity $I_0$ and column density $\tilde\rho$ may vary spatially, we introduce the \emph{intensity-weighted average} of an operator $X$ across an aperture of radius $a$  as $\expect{X}_I^a=P_0^{-1}\int_0^a 2\pi r I_0 X\,dr$ where $P_0=\int_0^\infty 2\pi r I_0\, dr$ is the total probe power at the atoms.
The polarimeter signal is therefore $P_+-P_- = 2 \kappa P_0 \expect{\varphi_z}_I^a$ where $\kappa$ is the optical transmission accounting for losses at optical elements and the quantum efficiency of the detector.

Measurement of the spin projection is limited by intensity fluctuations in the probe beam, which contains contributions from technical noise and photon shot-noise.
The RMS shot-noise across the measurement interval $\tau_f$ is
$P_n = \sqrt{\hbar\omega P_\text{det}/{\tau_f}}$ where $P_\text{det}=\kappa P_0 \expect{1}_I^a$ is the total power reaching the detector.

Following the approach of Smith \textit{et al.}~\cite{smith_faraday_2003}, we define the SNR as the ratio of inferred spin-projection to the associated RMS fluctuations caused by shot-noise in the probe,
\begin{equation}
\label{eq:snr_definition}
\text{SNR}\equiv\frac{|\expect{\op{F}_z}|}{\delta \op{F}_z} = \frac{|P_+-P_-|}{P_n} = \frac{2|\expect{\op{F}_z}|}{\xi_f}\sqrt{\frac{\tau_f \kappa P_0}{\hbar\omega\expect{1}_I^a}}\expect{\varphi_0}_I^a
.
\end{equation}
The factors $\expect{1}_I^a$ account for transmission through the aperture and $\expect{\varphi_0}_I^a$ is the spatially-averaged Faraday rotation angle.

The SNR can therefore be improved by increasing the optical power or detuning closer to resonance.
However, this increases off-resonant scattering of probe photons, which limits the possible duration of the measurement.
We therefore consider optimizing the SNR at a fixed scattering rate $\gamma_s$ by appropriate choice of $P_0$, and varying the detuning.

The scattering rate $\gamma_s$ for a given probe intensity $I_0$ is estimated with the Kramers-Heisenberg relation in App.~(\ref{app:scattering}).
Summing over all intermediate excited states within the fine-structure doublet gives $\gamma_s = \gamma_0 I_0/\xi_s^2$, where
\begin{align}
\gamma_0 & 
	= \frac{\omega^3 \alpha_0^2}{18\pi \epsilon_0^2 \hbar^3 c^4} 
\quad\text{and}\quad
\frac{1}{{\xi}_s^2}
	= \sum_{J'F'} \frac{\alpha^{(0)}_{J'F'}}{\alpha_0\Delta_{J'F'}^2}.
\label{eq:scatter_rate}
\end{align}
We term $\xi_s$ the ``incoherent weighted detuning'' accounting for different scattering rates on different transitions.

The scattering lifetime $\tau_s$ of the cloud is the inverse of the density-weighted average of scattering rates across the cloud,
\begin{equation}
\frac1{\tau_s} \equiv \int \frac{\gamma_s\tilde\rho\,dA}{N} = \frac{\gamma_0P_0}{N\xi_s^2}\expect{\tilde\rho}_I^\infty .
\end{equation}
Hence the SNR can be rewritten as
\begin{equation}
\label{eq:snr_general}
\text{SNR} = \sqrt{\kappa N}\frac{3\lambda}{\sqrt{2\pi}}\frac{\expect{\tilde\rho}_I^a}{\sqrt{\expect{1}_I^a \expect{\tilde\rho}_I^\infty}} \left\vert \frac{\xi_s}{\xi_f} \right\vert \sqrt{\frac{\tau_f}{\tau_s}}\expect{\op{F}_z} .
\end{equation}
This expression clearly separates the dependence of the SNR on geometry, atomic energy-level structure, and measurement timescales, allowing individual optimization of each.

\begin{figure}
    \centering
    \includegraphics[width=1.00\columnwidth]{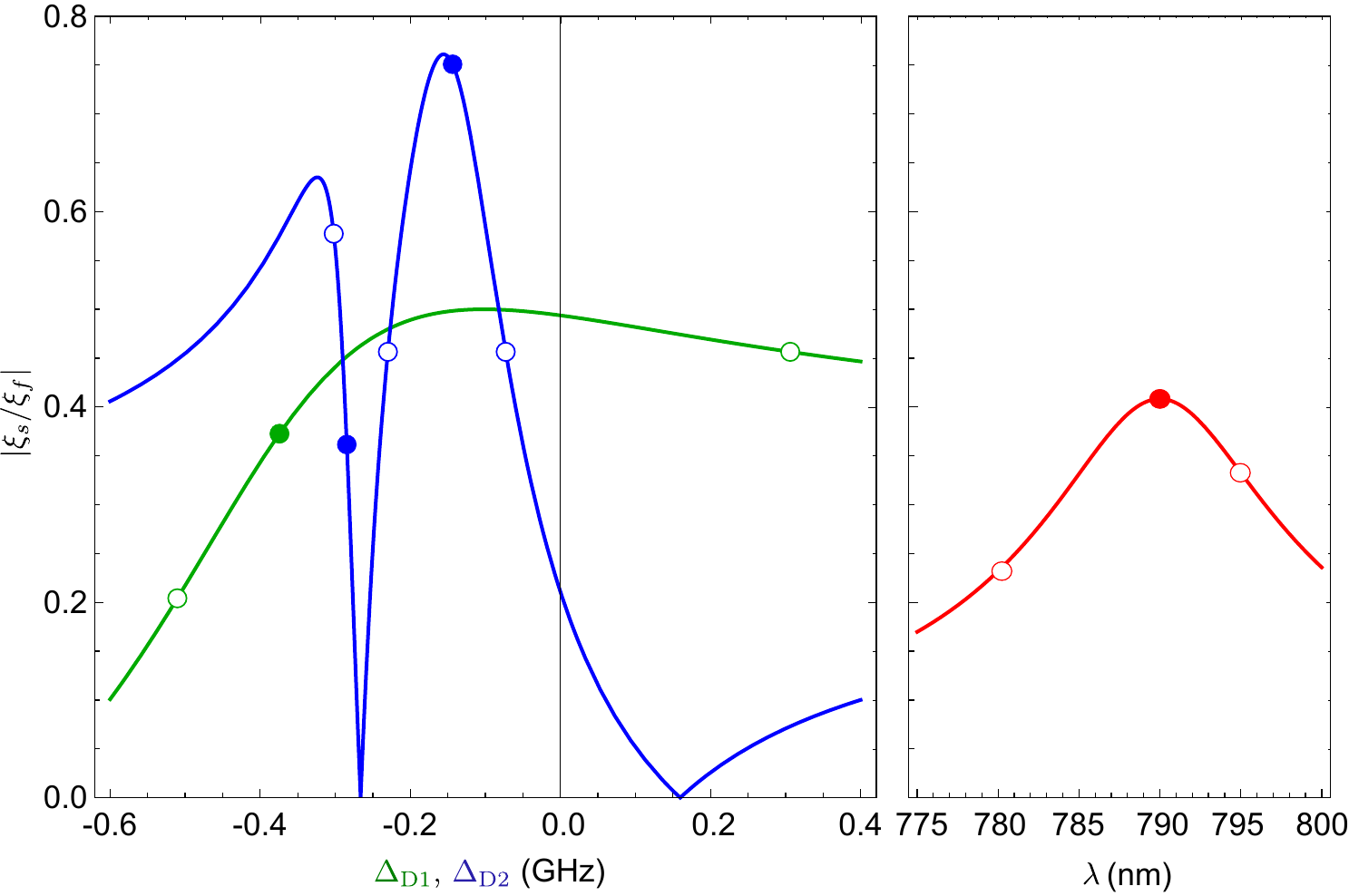}
    \caption{
        \label{fig:snr}
        Atomic component of the SNR for detunings near the D1 (green) and D2 (blue) lines (left) versus between the two lines (right).
                Open circles correspond to hyperfine- (left) or fine-structure (right) resonances; closed circles correspond to magic-zero wavelengths.
                Constructive (destructive) interference between the contributions from adjacent transitions leads to a maxima (zeros) in $|\xi_s/\xi_f|$ at $\Delta_\text{D2}/2\pi=\unit[-0.16\;(-0.27)]{GHz}$.
                Near-resonant effects (at detunings less than the natural linewidth) are not accounted for by \refeq{eq:Hint} or included in this plot.
    }
\end{figure}

If the probe intensity is constant across the detection aperture, the weighted averages can be easily evaluated to give
\begin{equation}
\label{eq:snr_uniform_I0}
\text{SNR} = N_a \sqrt{\kappa} \frac{3\lambda}{\sqrt{2\pi A}} \left\vert \frac{\xi_s}{\xi_f} \right\vert \sqrt{\frac{\tau_f}{\tau_s}} \expect{\op{F}_z} ,
\end{equation}
where $N_a = \int_0^a 2\pi r \tilde\rho \,dr$ is the number of atoms contributing to the measurement and $A=\pi a^2$ is the detection area.
The aperture size is easily optimized by maximizing $N_a/\sqrt{A}$ for a given cloud profile.
The optimum for a Gaussian cloud with $1/e^2$ radius $R$ is $a=0.79R$, whereas for a Thomas-Fermi cloud with radius $R$ it is $a=0.73R$.

The atomic dependence of the SNR as a function of detuning is captured in the ratio $\xi_s/\xi_f$.
We now consider a \Rb condensate in $F=1$ with probe detuning in one of three regimes: (I) near a single fine-structure line, (II) between the two lines, and (III) within the hyperfine structure of a single line.

\textit{Regime I:} For probe detuning near the D2 line ($|\Delta_\text{D2}| \ll |\Delta_\text{D1}|$) but far-detuned with respect to the excited-state hyperfine splitting, $\xi_f \approx 3\sqrt{2}\,\xi_s$, and
\begin{equation}
\label{eq:snr_d2}
\text{SNR}_\text{(D2)}
	= \frac{\lambda  N_a \sqrt{\kappa}}{4\pi a} \sqrt{\frac{\tau_f}{\tau_s}}.
\end{equation}
This agrees with~\cite{smith_faraday_2003}, which assumes a Gaussian cloud profile with optimum aperture.

\textit{Regime II:} In the far-detuned limit between the D1 and D2 lines, the dependence of the SNR is captured in the ratio
\begin{equation}
\frac{\xi_s}{\xi_f}
	\approx \frac{\Delta_\text{D1}^{-1} - \Delta_\text{D2}^{-1}}{3\sqrt{\Delta_\text{D1}^{-2} + 2\Delta_\text{D2}^{-2}}}
	= \frac{({2}/{\sqrt{3}})\omega_\text{fs}}{\sqrt{(6\Delta + \omega_\text{fs})^2 + 8\omega_\text{fs}^2}},
\label{eq:snr_magic}
\end{equation}
where $\Delta = \frac12(\Delta_\text{D1} + \Delta_\text{D2})$ is the detuning from the centre of the lines and $\omega_\text{fs}$ is the fine-structure splitting.
The Faraday contributions from each line add coherently, but the scattering contributions add in quadrature.

This expression is maximized at $\Delta = -\omega_\text{fs}/6$, giving $\lambda=\unit[790.0]{nm}$ and $\xi_s/\xi_f=\sqrt{2}/3$.
This detuning is precisely where the scalar polarizability vanishes (a ``magic-zero'' wavelength).
Measurement at this wavelength is fortuitously doubly-optimal in terms of both SNR and minimizing the trap perturbation discussed previously.

\textit{Regime III:} When detuning between hyperfine transitions of the D2 line (\reffig{fig:snr}), the SNR is maximized at $\Delta_\text{D2}/2\pi=-\unit[156]{MHz}$ with $|\xi_s/\xi_f|=0.76$, which does not correspond to a magic-zero wavelength.
There is therefore a trade-off between SNR and induced scalar light-shift perturbation.
Choosing the nearby magic-zero wavelength at $\Delta_\text{D2}/2\pi=-\unit[143]{MHz}$ eliminates this perturbation with a nominal decrease in SNR to $|\xi_s/\xi_f|=0.75$.

In comparing the above three regimes as candidates for a minimally-perturbative atom-light interface, Regime I is discounted as it has no proximate magic-zero wavelength.
In principle there appears to be little advantage between operating at magic-zero wavelengths in Regimes II and III as the $|\xi_s/\xi_f|$ ratios are similar.
However, the closer detuning in Regime III necessitates a $10^5$ reduction in probe power to achieve the same scattering rate as in Regime II, making continuous shot-noise limited photodetection far more challenging.
This motivates the choice of $\lambda=\unit[790.0]{nm}$ to perform continuous Faraday probing, which we exclusively consider hereafter.

\section{Apparatus}
\label{sec:apparatus}
Our spinor Bose-Einstein condensate (BEC) apparatus~\cite{wood_magnetic_2015} loads $3\times10^9$ \Rb atoms in a six-beam magneto-optical trap from a Zeeman slower.
The laser cooled atoms are optically pumped into the $\ket{F=1,m_F=-1}$ state before they are evaporatively cooled in different conservative potentials: first a hybrid magnetic optical dipole trap~\cite{lin_rapid_2009}, followed by a crossed-beam dipole trap formed using a $\unit[20]{W}$ fiber laser at $\unit[1064]{nm}$.
Care was taken to extinguish vector-light shifts from the dipole trapping light~\cite{wood_vls_2016}, which would otherwise result in a Zeeman-state dependent confining potential and premature dephasing of the collective condensate spin.  
The BEC is typically comprised of $~3\times10^5$ atoms, held in a harmonic potential with trapping frequencies $\times\unit[(35,60,80)]{Hz}$.
Three orthogonal sets of coil pairs generate magnetic fields of up to $\unit[20]{G}$, with individual control over the current in each coil enabling the generation of magnetic field gradients.
Conventional time-of-flight absorption imaging is routinely used for diagnostic purposes such as optimizing the cooling and trapping that precedes Faraday detection.
Alternatively, an absorption image of the cloud can be taken after it has been dispersively interrogated in-trap with the off-resonant probe to measure number loss and observe any changes in cloud structure.

The probe light is generated with a diode laser tuned to $\unit[790]{nm}$, with up to $\unit[16]{mW}$ reaching the science chamber.
The probe is tuned to the magic wavelength using a HighFinesse~\textsc{wsu-10} wavemeter with a short-term stability of $\unit[10]{MHz}$. 
Active frequency stabilization is not required due to insensitivity of the atom-light interaction so far from resonance, requiring only temperature stabilization.

Despite lasing at $\unit[790]{nm}$, the diode laser producing the Faraday beam has a broad amplified
spontaneous emission (ASE) background that spans $770$-–$\unit[810]{nm}$.
This background emission is weak, but contains a small component that is resonant with
the atomic transitions at $\unit[780]{nm}$ (D2 line) and $\unit[795]{nm}$ (D1 line), significantly decreasing the BEC lifetime.
An interference filter rotated for transmission at $\unit[790]{nm}$ blocks the resonant components increasing the $1/e$ lifetime from $\unit[29(6)]{ms}$ to $\unit[1.18(3)]{s}$ for $\unit[9.7(1)]{mW}$ of probe power.
Measurements of the BEC lifetime at various probe beam powers yield a scattering rate of $\unit[85(1)]{s^{-1}W^{-1}}$, consistent with \refeq{eq:scatter_rate}.
These measurements account for the dominant collisional loss mechanisms: three-body collisions between condensate atoms~\cite{soding_three-body_1999} and one-body collisions with background atoms in the vacuum.
The one-body limited lifetime in the absence of probe light is $\unit[35.0(7)]{s}$.

\begin{figure}
    \centering
    \includegraphics[width=1.00\columnwidth]{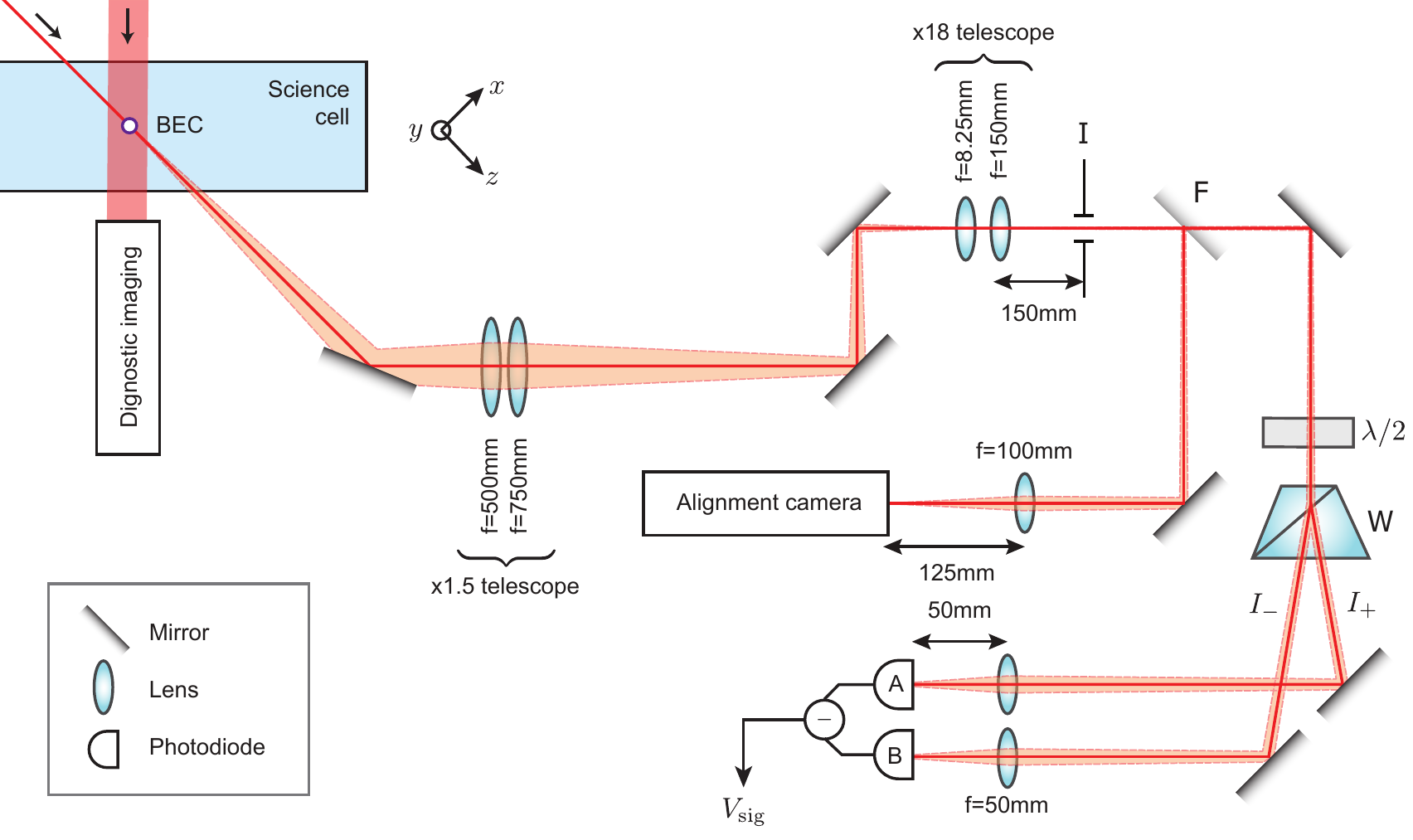}
    \caption{
        \label{fig:apparatus}
        Faraday atom-light interface: the probe is focused onto the condensate then magnified and re-imaged onto an iris to aperture the beam. 
        A flipper mirror (F) reflects the beam to either a camera or balanced polarimeter formed by a Wollaston prism (W) and dual-port differential photodetector (A-B).
		}
\end{figure}

The probe beam is focused to a $\unit[150]{\upmu m}$ $1/e^2$ diameter at the BEC (geometric mean Thomas-Fermi radius $\unit[19]{\upmu m}$) for near-uniform~\footnote{
	Despite the weak focus, residual inhomogeneity of the probe beam intensity still causes significant dipole forces via scalar and vector light shifts at wavelengths other than magic wavelength or elliptical polarizations.
} illumination of the atoms.
Limited optical access requires relay lenses to collect the transmitted probe light, which is magnified by a microscope objective and re-imaged onto a iris to block light not interacting with the BEC (\reffig{fig:apparatus}).
The Faraday probe beam, iris, and trapped condensate are co-aligned by coupling resonant light into the same beam path and performing in-trap absorption imaging with a machine vision CCD camera.

The polarization rotation is measured using a Wollaston prism placed upstream from a differential photodetector~\cite{hobbs_tears_1997} using large-area ($\unit[13]{mm}^2$) Hamamatsu~\textsc{s1223-01} photodiodes, chosen for their relatively low capacitance.
The detector has a transimpedance gain of $\unit[1]{V/mA}$ with an AC-coupled $100\times$ second-stage amplifier and a variable gain third-stage voltage amplifier to match the input voltage range of the data acquisition hardware.
The measured noise-equivalent power is $\unit[140]{\upmu W}$ in bandwidth $\unit[8]{MHz}$.
Optical losses and the iris result in $\unit[2]{mW}$ of probe light typically recorded on each photodiode.

The BEC is initially spin-polarized along the $+y$-axis in a uniform magnetic field of $B_y=\unit[1]{G}$, perpendicular to the propagation axis of the probe beam.
In this configuration, $\expect{\op{F}_z}=0$ and the Faraday probe experiences no polarization rotation.
The probe is switched on for $\unit[20]{ms}$ to ascertain the photon shot-noise, before tipping the spins into the $x$--$z$ plane with a resonant radiofrequency $\frac{\pi}2$-pulse, initiating Larmor precession and the generation of a Faraday signal at the Larmor frequency $f_L$.
The analog signal produced by the photodetector is digitized by a National Instruments~\textsc{pcie-6363} 16-bit DAC with a maximum acquisition rate of $2\unit{MS/s}$~\footnote{The Nyquist frequency is therefore $\unit[1]{MHz}$, limiting the maximum Larmor frequency and thus magnetic field strength to $\unit[1.4]{G}$ without under-sampling. We have subsequently incorporated an AlazarTech \textsc{ats9462} ditigizer ($16$-bit, $\unit[180]{MS/s}$) to acquire Faraday signals at higher magnetic fields; the maximum Larmor frequency is then limited by the bandwidth of the detector.}.

\section{Signal structure and optimization}
\label{sec:sig_process}
\begin{figure}
    \centering
    \includegraphics[width=1.00\columnwidth]{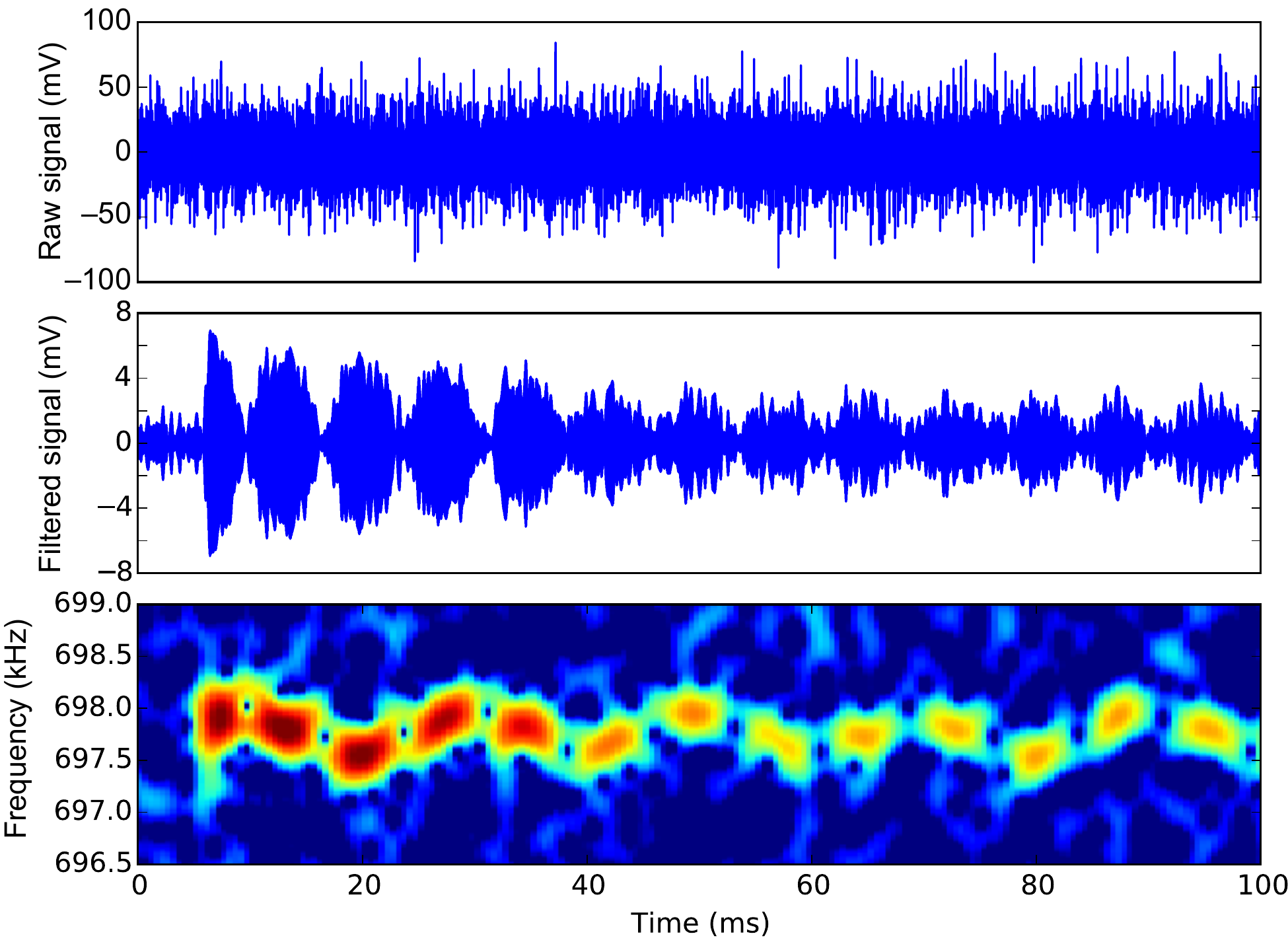}
    \caption{
        \label{fig:signal}
        Typical raw polarimeter signal (top) buried in noise, which is reduced by a $\unit[10]{kHz}$ band-pass filter to obtain the Faraday signal (middle) showing periodic revival.
                A spectrogram reveals both amplitude and frequency modulation of the Faraday signal (bottom).
                The average Larmor frequency is $f_L=\unit[697.8(4)]{kHz}$.}
\end{figure}

The raw measured polarimeter signal is dominated by photon shot-noise and low-frequency thermal fluctuations (\reffig{fig:signal}, top).
Applying a band-pass filter around the Larmor frequency reveals the Faraday signal (\reffig{fig:signal}, middle), which exhibits multiple decays and revivals within an overall free-induction decay (FID) envelope.

More structure is revealed using a short-time Fourier transform (STFT) algorithm, which divides the signal into overlapping `windows' and combines the magnitude of the Fourier transforms of each window into a spectrogram (\reffig{fig:signal}, bottom).
The Gabor limit defines a trade-off between frequency and temporal resolution.
The present analysis over-samples in both time and frequency; overlapping rectangular $\unit[5]{ms}$ windows are taken every $\unit[1]{ms}$ and zero-padded before Fourier transforming.
This smooths the spectrogram which clearly demonstrates both amplitude and frequency modulation, which is not clear from the bandpass-filtered time-domain signal.

The primary cause of amplitude modulation is the quadratic Zeeman effect (QZE), which cannot be neglected for strong bias fields and long interrogation times.
In the single-mode approximation, the spinor wavefunction is $(\sqrt{\rho_{-}} e^{i \Theta_{-}}, \sqrt{\rho_{0}} e^{i \Theta_{0}}, \sqrt{\rho_{+}} e^{i \Theta_{+}})$, where $\rho_i$ and $\Theta_i$ are the fractional populations and phases of the Zeeman sublevels.
Spin-polarized states with $\rho_+=\rho_-$ maximize $\expect{\op{F}_z}$, and therefore the induced Faraday signal (\refeq{eq:snr_general}).
The spin projection evolves under the mean-field Hamiltonian including spin-exchange as~\cite{zhang_coherent_2005}
\begin{equation}\begin{aligned}[c]
\label{eq:zhang1}
\expect{\op{F}_z} &= 2\sqrt{\rho_0(1-\rho_0)} \cos(\tfrac12\Theta) \cos(\omega_L t) , \\
\textstyle\partialD{\rho_0}t &= \tfrac{2c}{\hbar}(1-\rho_0)\sin\Theta , \\
\textstyle\partialD{\Theta}t &= -2q_z + \tfrac{2c}{\hbar}(1-2\rho_0)(1 + \cos\Theta) , \\
\omega_L &\equiv (E_{+1}-E_{-1})/2\hbar, \\
     q_z &\equiv (E_{+1}+E_{-1}-2E_0)/2\hbar, 
\end{aligned}\end{equation}
where $\Theta = \Theta_+ + \Theta_- - 2\Theta_0$ is the spinor phase, $E_m$ is the energy of the $\ket{F,m}$ state, $c\propto a_2-a_0$ is a spin-mixing coefficient, and $a_f$ are the elastic scattering lengths for two $F=1$ atoms colliding with total spin $f$.

\begin{figure}
    \centering
    \includegraphics[width=1.00\columnwidth]{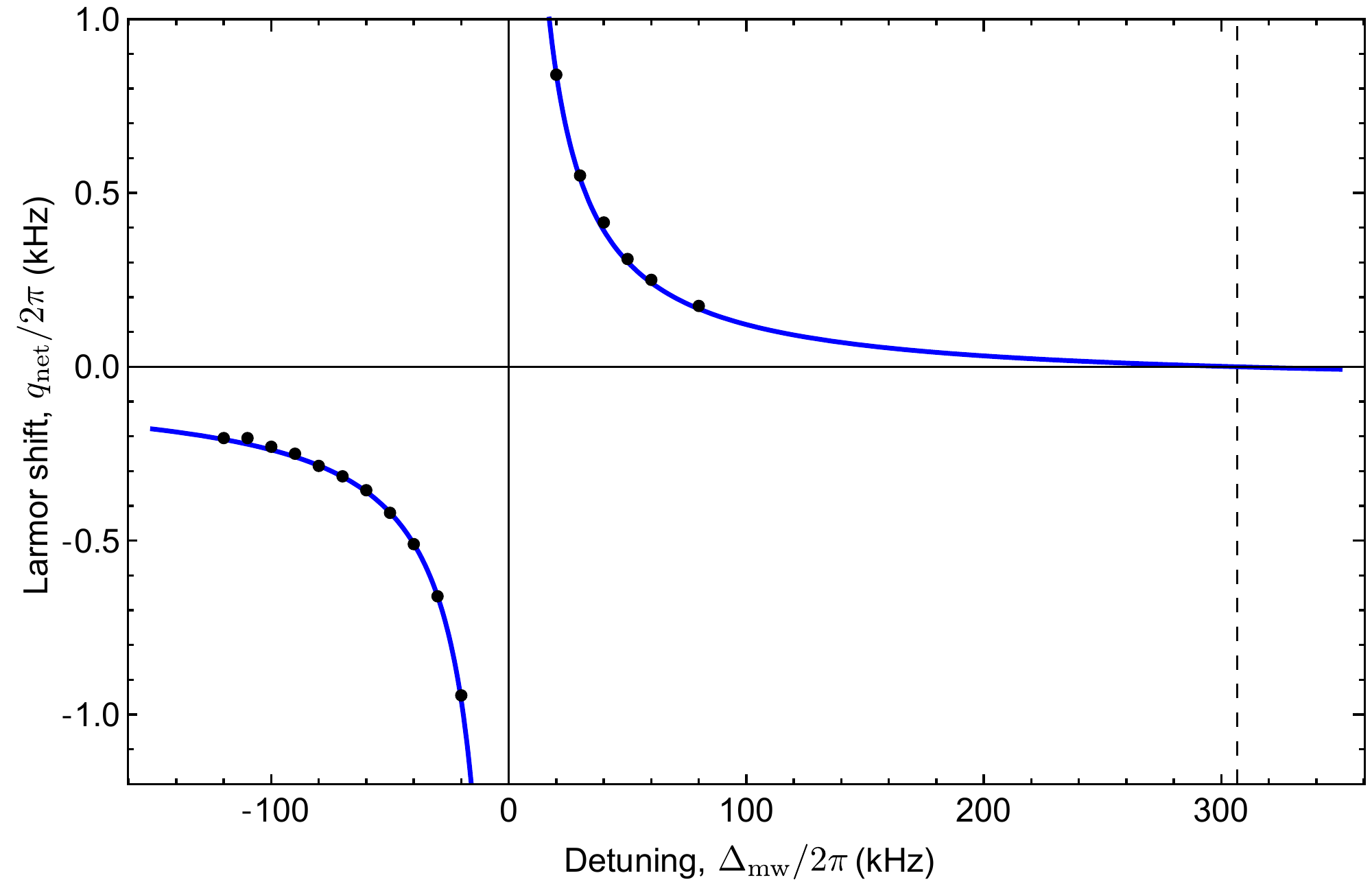}
    \caption{
        \label{fig:microwaves}
        Measured effective quadratic shift $q_\text{net} = q_z - q_\text{mw}$ in the presence of microwaves detuned by $\Delta_\text{mw}$ above the $\ket{F=1,m_F=0}\leftrightarrow\ket{F=2,m_F=0}$ clock transition at $f=\unit[6834682610.904]{Hz}$.
                The effective shift vanishes at $\Delta_\text{mw}/2\pi = \unit[307(2)]{kHz}$ (dashed line) and inferred microwave Rabi frequency is $\Omega_\text{mw}/2\pi=\unit[8.50(2)]{kHz}$.
        }
\end{figure}

This result demonstrates that the spin projection $\expect{\op{F}_z}$ and therefore the Faraday signal oscillates at $\omega_L$, which is linear in $B$ to $\mathcal{O}(B^3)$.
However, the signal is amplitude modulated directly through the cosine term, and indirectly through fluctuations in $\rho_0$ arising from spin-mixing.
When the QZE dominates ($q_z \gg c/\hbar$), the spinor phase winds linearly, $\Theta = 2 q_z t$, and fluctuations in $\rho_0$ are ``frozen out'' with $\rho_0=1/2$.
Consequently $\expect{\op{F}_z}$ and the Faraday signal is amplitude modulated at $q_z$~\footnote{
	If the spectrogram resolution is much finer than the quadratic splitting, $\Delta f < q_\text{net}/2\pi$, the amplitude modulation is instead resolved as two sidebands at $f_\pm = f_L \pm q_\text{net}/2\pi$.
}.

Nonlinear Zeeman shifts are a common problem in optical magnetometry, and are addressed by a variety of techniques~\cite{budker_book_2013}.
In our system, such shifts cause coherent spin evolution and do not limit measurement duration.
However, truly continuous magnetometry requires this amplitude modulation be suppressed to prevent $\expect{\op{F}_z}=0$.

This can be achieved at arbitrary magnetic field strengths by applying off-resonant microwave coupling~\cite{gerbier_resonant_2006} detuned from the $\ket{F=1,m_F=0}\leftrightarrow\ket{F=2,m_F=0}$ transition.
For microwaves with Rabi frequency $\Omega_\text{mw}$ and detuning $|\Delta_\text{mw}| \gg \Omega_\text{mw}$, the population of $F=2$ is minimal, and the induced quadratic shift is $q_\text{mw} \approx -\Omega_\text{mw}^2/4\Delta_\text{mw}$~\footnote{Note that imperfect linear polarization of the microwave source causes a correction to the induced quadratic shift from the other Zeeman substates.}.
Hence the QZE can be suppressed by appropriate choice of the microwave detuning (\reffig{fig:microwaves}), as $q_z$ is consistent with the Breit-Rabi equation~\cite{breit_rabi_1931}.

Amplitude modulation also arises from imperfect linear polarization of the probe beam causing evolution of the atomic spin state through the vector light shift.
Spatial variation of the probe intensity renders this an effective magnetic field gradient~\cite{wood_vls_2016}, which dephases the collective spin and limits the measurement time.
\begin{figure}
    \centering
    \includegraphics[width=1.00\columnwidth]{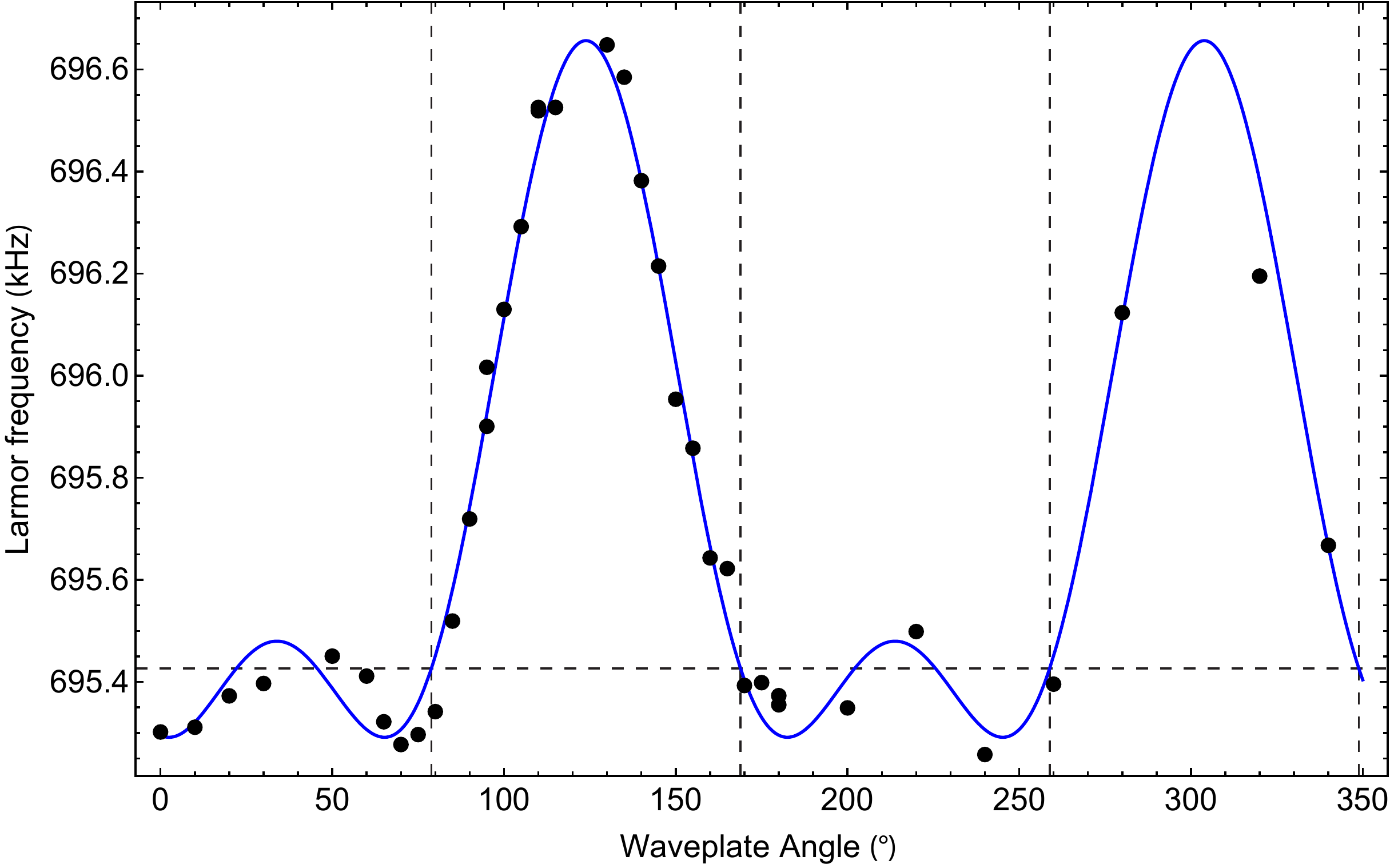}
    \caption{
        \label{fig:faraday_vls}
        Larmor frequency as a function of $\lambda/4$-waveplate angle ($\theta$), showing the shift induced by the VLS and the angles at which the VLS vanishes (dashed lines).
                The background field component is $B_z = \unit[19.6(8)]{mG}$, maximum VLS strength is $B_\text{vls}^{(0)}=\unit[43(1)]{mG}$ and $\theta_0=78.9(4)^\circ$.
        }
\end{figure}

This dephasing can be eliminated by using a uniform intensity probe, or making the polarization perfectly linear.
Although the polarization of the probe is purified using a Glan-Laser polarizer (extinction $10^5:1$) before the vacuum window, birefringence of optical elements after this polarizer causes an elliptically polarized probe at the atoms.
In the quasi-static approximation, \refeq{eq:H1} can be rewritten 
\begin{equation}
\H_z = \frac{\upmu_B g_F}{\hbar} B_\text{vls} \op{F}_z,
\end{equation}
where $B_\text{vls} \propto \expect{\op{S}_z}$ is the effective magnetic field in the $z$-direction induced by the probe beam ellipticity.

A quarter-waveplate at angle $\theta$ before the science cell enables control of the ellipticity within the cell through $B_\text{vls} = B_\text{vls}^{(0)} \sin(2(\theta-\theta_0))$, where $B_\text{vls}^{(0)}$ is the VLS for a circularly polarized probe and $\theta_0$ is the waveplate angle at which the polarization is linear at the atoms.
$B_\text{vls}$ introduces a small shift to the Larmor frequency,
\begin{equation}
\omega_L = \frac{\upmu_B g_F}{\hbar} \sqrt{B_y^2 + (B_{z}+B_\text{vls})^2} ,
\end{equation}
where $B_y$ is the dominant bias field component and $B_{z}$ is the (small) background field component in the $z$-direction.
The measurement is sufficiently sensitive that the VLS component can be extracted as a function of $\theta$, and hence $\theta_0$ determined (\reffig{fig:faraday_vls}).
\begin{figure}
    \centering
    \includegraphics[width=1.00\columnwidth]{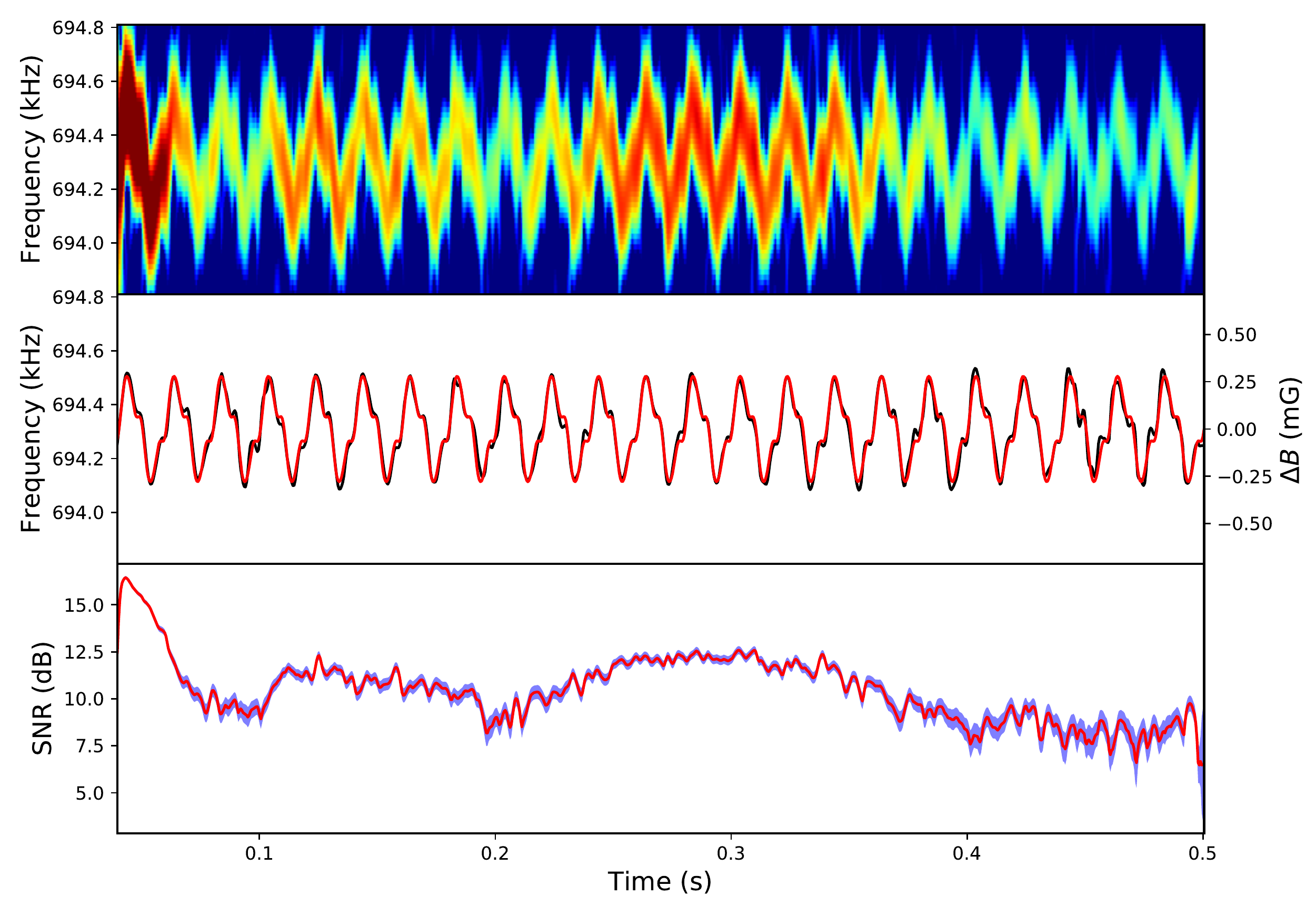}
    \caption{
        \label{fig:powerline}
        Spectrogram with QZE eliminated (top) enables continuous measurement of the Larmor frequency (middle) for time-dependent magnetometry.
        Larmor frequency from peak fitting in each spectrogram window (middle, black); this frequency modulation is modeled by the first three harmonics of the power line (red).
        Measuring the peak SNR at each spectrogram window (bottom) clearly shows residual amplitude modulation signal, which we attribute to ambient magnetic field gradients.}
\end{figure}

\section{Microscale atomic magnetometry at high temporal resolution}
With the primary source of amplitude modulation eliminated, we achieve continuous magnetometry (\reffig{fig:powerline}) revealing time-dependent magnetic field fluctuations integrated across the $(\unit[30]{\upmu m})^3$ sensing volume of the atomic condensate.
This time-dependence is manifest as frequency modulation of the Faraday signal, in this instance containing odd harmonics of the $\unit[50]{Hz}$ power-line frequency due to nearby electronic equipment.
\begin{figure*}[ht]
    \centering
    \includegraphics[width=\textwidth]{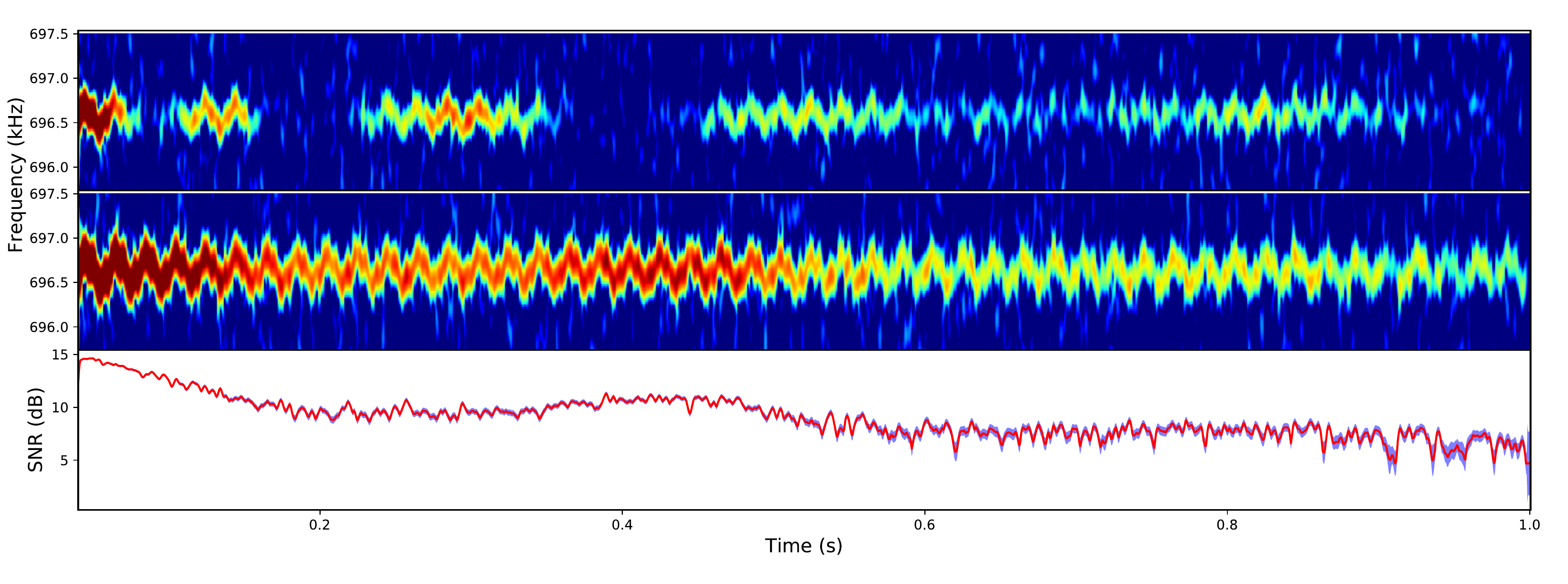}
    \caption{
        \label{fig:Bzgrad_sig}
        Measured Faraday signal without (top) and with (middle) cancellation of both VLS and background magnetic field gradients showing (bottom) gradual decay of the SNR over a $\unit[1]{s}$ measurement duration.
        (Top) The quadratic Zeeman shift has been canceled using microwave dressing (\refsec{sec:sig_process}) but magnetic field gradients due to vector light shifts and background fields remain.
        We contend that the irregular amplitude modulation is due to the rich dynamical interplay between these gradients, spin-mixing dynamics of the condensate, and the concomitant breakdown of the single-mode approximation.
        (Middle) Ensuring linear polarization of the probe light and applying a counter-gradient results in reduced dephasing and achieves long-lived continuous readout of the spin projection.
        (Bottom) Prolonged SNR exceeding unity beyond $\unit[1]{s}$.
        Off-diagonal gradients $\lesssim \unit[1]{mG/cm}$ which were not canceled result in coherent decay and revival of the cloud, causing the residual amplitude modulation. 
        }
\end{figure*}

The peak-to-peak variation of this pickup is $\Delta B_\text{ac} \approx \unit[558(4)]{\upmu G}$.
We resolve the harmonic components of the time-varying magnetic field, finding amplitudes of $162(1)$, $46(1)$, and $\unit[7(1)]{\upmu G}$  at $50$, $150$, and $\unit[250]{Hz}$, respectively.

The spectrogram (\reffig{fig:powerline}, top) shows evidence of residual amplitude modulation, which is clarified by plotting the peak amplitude in each time window (\reffig{fig:powerline}, bottom).
The remaining free induction decay of the signal is caused by the small gradients in real magnetic fields, as opposed to light-induced gradients nulled in Sec.~\ref{sec:sig_process}.
A spherical cloud of radius $R$ has dephasing timescale $\tau_D \sim \pi/2\gamma b R$ where $b=\sum_i \partial B_y / \partial x_i$ in a $y$-oriented magnetic field.
The ambient gradients in our apparatus were found to be $\sim \unit[10]{mG/cm}$, measured independently using a differential Ramsey interferometer~\cite{wood_magnetic_2015}, resulting in a dephasing time of $\tau_D \sim \unit[20]{ms}$.
To achieve continuous magnetometry on the timescale of seconds, we apply a counter-gradient $\partial B_y/dy$ using asymmetric currents in the bias magnetic field coils.
The resulting free-induction decay (\reffig{fig:Bzgrad_sig}) shows an SNR exceeding unity beyond $\unit[1]{s}$.
The gradient was nulled to within $b \lesssim \unit[1]{mG/cm}$ ($\tau_D \sim \unit[200]{ms}$), limited by off-diagonal gradients (e.g. $\partial B_y/dz$).
These can only be canceled using a specific arrangement of gradient coils~\cite{wood_magnetic_2015} not available in this experiment.

We determine the initial SNR to be $\unit[16.3(2)]{dB}$ from the spectrogram data shown in \reffig{fig:powerline}~\footnote{We average the shot-noise over a $\unit[20]{ms}$ interval prior to tipping the spins across a $\unit[2]{kHz}$ frequency band around the mean Larmor frequency.}.
For a Thomas-Fermi density profile and Gaussian probe beam, \refeq{eq:snr_general} predicts a shot-noise limited SNR of $\unit[18.3]{dB}$ provided $N=3\times10^5$ atoms, optical transmission $\kappa=0.2$, aperture radius $a=\unit[38]{\upmu m}$ equal to twice the geometric mean Thomas-Fermi radius, $|\xi_s/\xi_f|=\sqrt{2}/3$, and scattering lifetime $\tau_s=\unit[1.2]{s}$. 
By recomputing spectrograms with different window length $\tau_f$, we confirmed that the measured SNR scales with $\sqrt{\tau_f}$ as predicted.

The ratio of the atomic standard quantum limit (SQL) to the photon shot-noise limit in \refeq{eq:snr_definition} can be expressed as $(\delta \op{F}_z)_{\text{SQL}}/\delta \op{F}_z = \sqrt{\kappa \sigma_0 \tilde{\rho} \tau_f / (2\tau_s)}$, where $\sigma_0$ is the resonant photon scattering cross-section~\cite{romalis_quantum_2013}.
Our measurement is dominated by photon shot-noise for $\tau_f < \unit[12.5]{ms}$, justifying the analysis in \refsec{ssec:snr_optimization}.
For our parameters, $(\delta \op{F}_z)_{\text{SQL}} = 0.63\delta \op{F}_z$, which decreases the predicted SNR by $\unit[0.7]{dB}$.

The SNR defined in \refeq{eq:snr_definition} is the reciprocal of the uncertainty in the Larmor phase of the collective spin $1/\delta \phi$, and can thus be used to estimate the magnetic field sensitivity per unit bandwidth $\delta B \sqrt{T} = 1/(\gamma \text{SNR} \sqrt{\tau_f})$, where $\gamma$ is the gyromagnetic ratio.
For these data ($\tau_f = \unit[5]{ms}$), we estimate the photon shot-noise limited field sensitivity to be $\delta B \sqrt{T}=\unit[7]{pT/\sqrt{Hz}}$, slightly below the experimental value of $\unit[10]{pT/\sqrt{Hz}}$ inferred from the standard error of the fitted Larmor frequency in a given spectrogram window (\reffig{fig:powerline}, middle).

\section{Concluding remarks}
\label{sec:conclusion}
In conclusion, we have demonstrated continuous Faraday measurement of a condensed spinor gas absent scalar and vector light shifts.
We evaluated the shot-noise limited signal-to-noise ratio for a given scattering rate, motivating the use of a bright, linearly polarized probe at $\lambda=\unit[790]{nm}$ to realize a minimally-perturbative atom-light interface.

Spectrogram analysis revealed the detail inherent in the continuous Faraday signal, making plain the quadratic Zeeman effect, vector light-shifts, and gradient-induced dephasing as amplitude modulations of the Larmor carrier.
We demonstrated how each of these can be canceled in turn, enhancing the contiguous measurement interval without dead time.
The resulting long interrogation times of $\sim\unit[1]{s}$ enable either close determination of the mean Larmor frequency for precision magnetometry near DC, or observation of time-dependent magnetic fields (manifest as frequency modulation of the Larmor carrier), in accordance with the Gabor limit.
Single-shot acquisition of $\unit[1]{million}$ polarimetry measurements resolved the amplitude of these low-frequency fluctuations to $\unit[1]{\upmu G}$ in $\unit[5]{ms}$ intervals, allowing separate harmonic components of parasitic field noise to be identified.

The minimally perturbative nature of the Faraday probe is ideal for quantum state estimation in cold atomic ensembles~\cite{smith_efficient_2006,riofrio_quantum_2011}.
In this work we applied microwave control to null the quadratic Zeeman shift and minimize amplitude modulation of the Faraday signal.
The quadratic Zeeman shift breaks the rotational symmetry of the spin degree of freedom, which is necessary to estimate density matrices of spins $> 1/2$.
To date quantum state estimators have exploited a fixed quadratic shift borne of a constant probe-laser tensor light shift.
Modulating the microwave dressing \emph{during} the measurement will permit pulsed, time-reversible tomographic state reconstruction.

Proposals to apply Faraday quantum non-demolition measurements as momentum-selective probes of strongly-correlated quantum gases~\cite{eckert_quantum_2008} may founder if the standing wave scalar light shifts (optical lattice) of the probe beam confounds the measurand.
Implementing these probes at a magic-zero wavelength obviates this impediment.

In precision magnetometry applications, the long measurement times we demonstrated open the possibility of single-shot 3-axis vector magnetometry by adiabatically rotating the magnetic field bias direction during the measurement.

Real-time data processing of the Faraday signal may be used for closed-loop control of magnetic fields by feedback to compensation coils, for example to suppress fluctuations as required by experiments preparing delicate spin-entangled many-body states.

\appendix
\section{Interaction coefficients}
\label{app:coeff}
The polarizability coefficients for the $\ket{JF}\rightarrow\ket{J'F'}$ transition are~\cite{stockton_continuous_2007}
\begin{align}
\alpha^{(0)}_{J'F'} &= \alpha_{JF}^{J'F'}\left(\delta_{F-1}^{F'} + \delta_F^{F'} + \delta_{F+1}^{F'} \right) ,\\
\alpha^{(1)}_{J'F'} &= \alpha_{JF}^{J'F'}\left(-\tfrac{1}{F}\delta_{F-1}^{F'} - \tfrac{1}{F(F+1)}\delta_F^{F'}  + \tfrac{1}{F'}\delta_{F+1}^{F'} \right) ,  \\
\alpha^{(2)}_{J'F'} &= \tfrac{\alpha_{JF}^{J'F'}}{2F'+1}\left(\tfrac{1}{F}\delta_{F-1}^{F'} - \tfrac{2F+1}{F(F+1)}\delta_F^{F'} + \tfrac{1}{F'}\delta_{F+1}^{F'} \right) , \\
\alpha^{J'F'}_{JF} &= \alpha_0 (2F'+1)(2J'+1)\left\vert \sixj{1}{J}{J'}{I_s}{F'}{F} \right\vert^2 , \label{eq:alpha_JF}\\
\alpha_{0} &
	= \frac{3\epsilon_0\hbar\lambda_{J'}^3\Gamma_{J'}}{8\pi^2},
\end{align}
where $\delta_m^n$ is the Kronecker delta, $I_s$ is the nuclear isospin, $\lambda_{J'}$ is the resonant wavelength and $\Gamma_{J'}$ is the natural linewidth of the transition.

The \emph{polarizability constant} $\alpha_0$ is independent of $J'$ as $\lambda_\text{D1}^3\Gamma_\text{D1} = \lambda_\text{D2}^3\Gamma_\text{D2}$.
This can be seen by relating $\Gamma_{J'}$ to the reduced dipole element {$\expect{J||\vec{d}||J'}$}, expanding in terms of {$\expect{L||\vec{d}||L'}$} and evaluating the associated Wigner-$6j$.

\section{Derivation of scattering rate}
\label{app:scattering}
In the far-detuned limit, second-order perturbation theory predicts that an atom in state $\ket{a}$ can transition to state $\ket{b}$ through absorption and emission of a photon via a state $\ket{j}$.
For scattering into solid angle $d\Omega$, the Kramers-Heisenberg relation gives the scattering rate as~\cite{martin_quantum_2013}
\begin{widetext}
\begin{equation}
\label{eq:kramheis}
\frac{d\gamma_{a\rightarrow b}}{d\Omega} = 
	\frac{I_0 \omega_\text{sc}^3}{(4\pi \epsilon_0)^2\hbar^3 c^4}
	\left\vert \sum_{\ket{j}}
	\frac{\expect{b|\vec\epsilon_\text{sc}\cdot\op{\vec{d}}|j}\expect{j|\vec\epsilon\cdot\op{\vec{d}}|a}}{\omega_{ja}-\omega} +
	\frac{\expect{b|\vec\epsilon\cdot\op{\vec{d}}|j}\expect{j|\vec\epsilon_\text{sc}\cdot\op{\vec{d}}|a}}{\omega_{ja}+\omega_\text{sc}}
\right\vert^2,
\end{equation}
\end{widetext}
where $\hbar\omega_{ja}=E_j - E_a > 0$ is the energy difference between $\ket{a}$ and $\ket{j}$, $I_0$ is the probe intensity, $\omega$ the probe frequency, $\vec\epsilon$ the polarization vector, and the subscript ``sc'' denoting the scattered photon.

Taking the probe as linearly polarized, integrating over all emission solid angles $d\Omega$, summing over the scattered photon polarizations $q$, and all possible final atomic states $\ket{b}$, the total scattering rate out of state $\ket{a}$ is
\begin{equation}
\gamma =
	\frac{I_0}{6\pi\epsilon_0^2\hbar^3c^4} \!\! \sum_{\ket{b}} (\omega-\omega_{ba})^3 \sum_q\left\vert
	\sum_{\ket{j}} \frac{\expect{b|\op{d}_{q}|j}\expect{j|\op{d}_0|a}}{\omega_{ja}-\omega}
\right\vert^2 .
\end{equation}

Applying selection rules, the $q=0$ terms require $F''=F$ and $m_F''=m_F$, so the process is Rayleigh scattering.
Whereas for $q=\pm1$, $m_F''\neq m_F$ and the process is Raman scattering.
Taking $|\omega_{ba}| \ll \omega$ and $\gamma_0 = \omega^3\alpha_0^2/{18 \pi \epsilon_0^2 \hbar^3 c^4}$, 
the associated scattering rates for the two processes are
\begin{align}
\gamma^{(q=0)} &= \frac{I_0 \gamma_0}3 \left(\sum_{J'F'} \frac{\alpha_{J'F'}^{(0)}}{\alpha_{0}\Delta_{J'F'}}\right)^2 , \\
\gamma^{(q=1)} + \gamma^{(q=-1)} &= 6 I_0 \gamma_0 \left(\sum_{J'F'} \frac{\alpha_{J'F'}^{(1)}}{\alpha_{0}\Delta_{J'F'}}\right)^2 .
\end{align}
This is consistent with a Kramers-Kronig interpretation of the scalar and vector Hamiltonians.

In our system, the total scattering rate can then be calculated using properties of the Wigner-$6j$ symbols as
\begin{align}
\label{eq:scatter_all}
\gamma 
	= {I_0 \gamma_0} \sum_{J'F'}\frac{\alpha_{J'F'}^{(0)}}{\alpha_{0}\Delta_{J'F'}^2}.
\end{align}

\bibliography{faraday}

\end{document}

%% file: preamble.tex
\usepackage{graphicx}
\usepackage{color}
\usepackage{amsmath,amssymb}
\usepackage{mathrsfs}
\usepackage{bm}
\usepackage{upgreek}
\usepackage{xspace}
\usepackage{units}
\usepackage[colorlinks,urlcolor=blue,citecolor=blue,linkcolor=blue]{hyperref}

\newcommand{\Rb}{$^{87}$Rb\xspace}

\newcommand{\reffig}[1]{\mbox{Fig.~\ref{#1}}}
\newcommand{\refeq}[1]{\mbox{Eq.~(\ref{#1})}}
\newcommand{\refsec}[1]{\mbox{Sec.~(\ref{#1})}}

\newcommand{\partialD}[2]{\frac{\partial #1}{\partial #2}}
\newcommand{\expect}[1]{\langle #1 \rangle}

\newcommand{\ket}[1]{\vert #1 \rangle \xspace}
\renewcommand{\H}{\ensuremath\op{\mathscr{H}}}
\renewcommand\vec[1]{\ensuremath\bm{#1}}
\newcommand*\op[1]{\ensuremath\hat{#1}}

\newcommand{\sixj}[6]{ \begin{Bmatrix}
  #1 & #2 & #3 \\
  #4 & #5 & #6 
 \end{Bmatrix}}